\documentclass[aps,twocolumn,superscriptaddress]{revtex4}
\usepackage{amsfonts,amssymb,amsmath}            
\usepackage{graphics,graphicx,epsfig}            

\newcommand{\identity}{\openone}

\newcommand{\proj}[1]{\mbox{$|#1\rangle \!\langle #1 |$}}

\newcommand{\half}{\mbox{$\textstyle \frac{1}{2}$}}
\newcommand{\ket}[1]{\left | #1\right \rangle}
\newcommand{\bra}[1]{\left \langle #1\right |}

\begin{document}

\title{Reproducing spin lattice models in strongly coupled
atom-cavity systems}

\author{Alastair \surname{Kay}}
\affiliation{Centre for Quantum Computation,
             DAMTP,
             Centre for Mathematical Sciences,
             University of Cambridge,
             Wilberforce Road,
             Cambridge CB3 0WA, UK}
\author{Dimitris G. \surname{Angelakis}}
\affiliation{Science Department, Technical University of Crete,
Chania, Crete, Greece, 73100} \affiliation{ Centre for Quantum
Technologies, National University of Singapore, 2 Science Drive 3,
Singapore 117543}

\date{\today}

\begin{abstract}
In an array of coupled cavities where the cavities are doped with an
atomic V-system, and the two excited levels couple to cavity photons
of different polarizations, we show how to construct various spin
models employed in characterizing phenomena in condensed matter
physics, such as the spin-$1/2$ Ising, XX, Heisenberg, and XXZ
models. The ability to construct networks of arbitrary geometry also
allows for the simulation of topological effects. By tuning the
number of excitations present, the dimension of the spin to be
simulated can be controlled, and mixtures of different spin types
produced. The facility of single-site addressing, the use of only
the natural hopping photon dynamics without external fields, and the
recent experimental advances towards strong coupling, makes the
prospect of using these arrays as efficient quantum simulators
promising.
\end{abstract}
\maketitle


{\em Introduction:} The burgeoning field of quantum computation
promises much to the science and technology community. While the
ability to factor large numbers efficiently may still be some way
off, the advances and potential applications brought along with the
understanding and control of quantum processes, from beautiful
manipulations on minute systems \cite{walborn} through to coherent
many-body operations \cite{Greiner:2003a}, cannot be underestimated.
One of the first such applications is likely to be the simulation of
one quantum system with another, more easily manipulated, quantum
system. The most general results have been expressed by showing how
to simulate one Hamiltonian with another with the help of a series
of extremely fast single-qubit rotations, breaking the evolution
down into a sequence of stroboscopic pulses which approximate the
desired evolution \cite{Dodd}, which is known as a Trotter decomposition. However, in physical systems such as
optical lattices and ion traps, we possess much more direct ways of
simulating a variety of different systems, merely by adjusting
periodic potentials using, for example, globally applied lasers 
\cite{intro_optical}, making such simulations feasible with current
technology.

Of particular interest are models of the form
\begin{equation}
H=\sum_i{\vec B}\cdot{\vec \sigma}_i+\sum_{\langle i,j\rangle}\lambda_zZ_iZ_j+
\lambda_xX_iX_j+\lambda_yY_iY_j, \label{eqn:gen}
\end{equation}
where $\langle i,j\rangle$ denotes all nearest-neighbour pairs on a
lattice of a particular geometry (typically, a 1D chain, or 2D
square lattice) and ${\vec \sigma}$ is the vector of Pauli matrices
$X$, $Y$ and $Z$. There are a number of special cases which are
commonly examined. For example, the Ising model ($\lambda_z\neq 0$)
in a transverse magnetic field ($B_x\neq 0$) is a simple
one-dimensional model which exhibits critical properties. Others
include the XX ($\lambda_x=\lambda_y$ and $\lambda_z=0$), Heisenberg
($\lambda_x=\lambda_y=\lambda_z$) and XXZ
($\lambda_x=\lambda_y\neq\lambda_z$). In two dimensional lattices,
such as the hexagonal lattice, simple topological models arise. One
possible test-bed for these ideas is an optical lattice setup where
the natural Bose-Hubbard Hamiltonian can be manipulated to produce
these topological, critical and other effects 
\cite{DDL03a,carollo-2005-95}. In addition, they are capable of
creating three-body terms and chiral interactions 
\cite{pachos-2004-70}.

Coupled cavities arrays (CCAs) have been initially proposed for the
implementation of quantum gates \cite{ASYE_PLA_07}. Recently,
intense interest has arisen from the demonstration that a
polaritonic Mott transition  and a Bose-Hubbard interaction can be
generated in these structures 
\cite{angelakis-2006,hartmann-2006-2,greentree-2006}. In the same
work  it was shown that the Mott state could be mapped directly to a
spin XX model \cite{angelakis-2006}. These papers lead to a plethora
of studies on various properties of CCAs in the direction of many
body simulations \cite{rest}, quantum computation \cite{AK_NJP_08} and
production of photonic entanglement \cite{entanglement}. The study of CCAs provides a theoretical framework that can be implemented using a
variety of technologies such as photonic crystals, toroidal
microcavities and superconducting qubits \cite{pbgs,supercond,toroid}.
Thus, the aforementioned results are not bound to a specific physical system.

In this paper, the aim is to extend this theoretical framework by restricting to the on-resonance, strong
coupling, case and examining how one might enrich the simulated model by
incorporating more complex atomic structures within the dopants, and
by utilising photons of differing polarisations; the goal being to
achieve as much of the generalised model described in
Eqn.~(\ref{eqn:gen}) as possible without resorting to a Trotter decomposition, which imposes additional experimental difficulties. While such decompositions are applicable to the original CCA proposals \cite{angelakis-2006,hartmann-2006-2,greentree-2006}, a proposal implementing similar models
through the rapid switching of a number of off-resonant time
dependent optical fields followed up by a Trotter expansion has
recently been proposed \cite{martin-spins}. Coupled cavity arrays are capable of
single-qubit addressing, so the corresponding local magnetic fields
${\vec B}$ in a spin model simulation are readily achieved. The key to
creating the desired $ZZ$ (which was absent in initial proposals \cite{angelakis-2006,hartmann-2006-2,greentree-2006}) interaction is by suitably selecting the
degeneracy of the energy levels of the dopant atoms. We show how an
atomic V-system is capable of achieving this. This will provide the additional benefit that 
simply by tuning the number of excitations in the system, a large range of different, higher dimensional, spin
models can be simulated. The possibility of
simulating high dimensional spins in the presence of strong dissipation using
constant external fields is also currently being
examined \cite{jaeyoon_spins}. In the present work, for the case of
small dissipation, we present a simpler scheme utilizing just the
natural photon hopping dynamics of CCAs, and no time dependent
external fields or detunings.

\begin{figure}
\begin{center}
\includegraphics[width=0.35\textwidth]{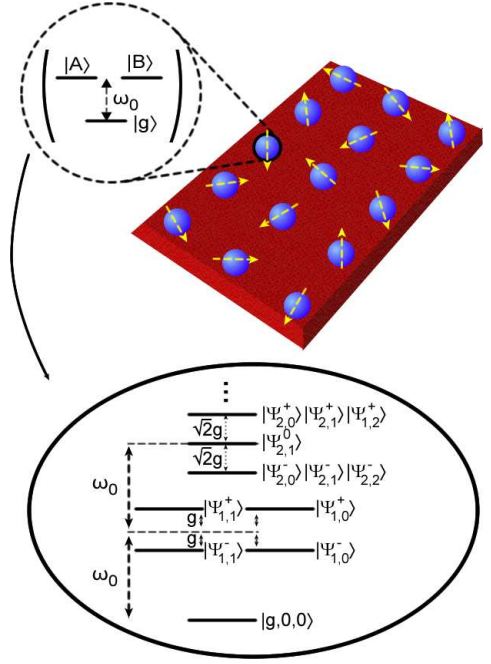}
\end{center}
\vspace{-0.5cm}
\caption{The atomic V-system on resonance with a cavity. There are
two orthogonal photon types, $a^\dagger$ and $b^\dagger$, which only
cause transitions to a single level ($A$ or $B$) from the ground
state. This gives rise to a non-linear internal structure on each site.}\label{fig:system}
\vspace{-0.5cm}
\end{figure}

{\em Atomic V System:} We start by considering an array of cavities,
placed on the vertices of an arbitrary lattice (typically, we
consider a regular lattice such as a 1D chain or 2D plane). Each
cavity is doped with a single system (which we refer to as an atom),
whose energy level structure is that of a ground state, $\ket{g}$
and two degenerate excited states $\ket{A}$ and $\ket{B}$, depicted
in Fig.~\ref{fig:system}. Within each lattice site, the Hamiltonian
takes the form
\begin{eqnarray}
H_{int}=&&\omega_0\left(aa^\dagger+bb^\dagger+\proj{A}+\proj{B}\right) \nonumber\\
&&\Delta_A\proj{A}+\Delta_B\proj{B} \nonumber\\
&& +g\left(\ket{A}\bra{g}\otimes a+\ket{g}\bra{A}\otimes
a^\dagger \right) \nonumber\\
&&+g\left(\ket{B}\bra{g}\otimes
b+\ket{g}\bra{B}\otimes b^\dagger\right),   \nonumber
\end{eqnarray}
where $a^\dagger$ and $b^\dagger$ create photons of orthogonal
polarisations, and are those responsible for promoting the ground
state of the atom to the excited states $\ket{A}$ and
$\ket{B}$ respectively. Henceforth, we assume that the atomic levels and
the cavity are on resonance (i.e.~the characteristic frequency of
the cavity is equal to the frequency of the atomic transitions of
the ground state to the excited states; $\Delta_A=\Delta_B=0$). The strength
$g$ represents the strength of the coupling between the cavity and
the atom.

In the basis
$\ket{\psi,N_A,N_B}$, we can calculate that the (unnormalised) on-site eigenvectors are
\begin{eqnarray}
\ket{\Psi_{S,n}^0}&=&\sqrt{S-n}\ket{A,n-1,S-n}-\sqrt{n}\ket{B,n,S-n-1}    \nonumber\\
\ket{\Psi_{S,n}^{\pm}}&=&\sqrt{n}\ket{A,n-1,S-n}+\sqrt{S-n}\ket{B,n,S-n-1}    \nonumber\\
&&\pm\sqrt{S}\ket{g,n,S-n}  \nonumber
\end{eqnarray}
with energies $S\omega_0$ and $S\omega_0\pm g\sqrt{S}$
respectively (see Fig. \ref{fig:system}).  $N_A$ and $N_B$ are the number of $a$ and $b$ photons
in the cavity, and $\psi$ is the state of the atom. Here, $n$ is an integer index ($0$ to $S$) which enumerates the basis within the manifold containing $S$ excitations.

Let us assume that we are working at unit filling fraction, so we
expect one excitation per lattice site, meaning that only the
states
\begin{eqnarray}
\ket{0}&=&(\ket{A,0,0}-\ket{g,1,0})/\sqrt{2}    \nonumber\\
\ket{1}&=&(\ket{B,0,0}-\ket{g,0,1})/\sqrt{2}    \nonumber
\end{eqnarray}
are populated. This arises from the observation that there is an
energy penalty of $U=(2-\sqrt{2})g$ for moving from one excitation per lattice site to having two
excitations in one site, and none in another.

\begin{figure}
\begin{center}
\includegraphics[width=0.4\textwidth]{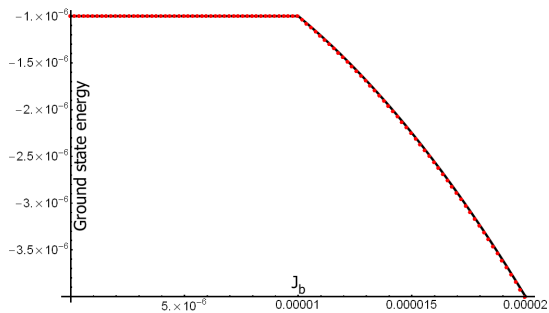}
\end{center}
\vspace{-0.5cm}
\caption{A comparison of the ground state energy between a simulation of the full system (dots) and the prediction from perturbation theory (solid line) for 4 cavities, doped with an average of 1 excitation per site. The chosen parameters are $g=10^{-3}$, $J_a=10^{-5}$. A phase transition occurs at $J_a=J_b$ between the $\ket{0}^{\otimes 4}$ and $\ket{0}^{\otimes 4}$ ground states. The energies have been scaled to remove the shift of $4\omega_0-4g$. To observe other phases, such as the one in the XXZ model requires compensation of the $B_z$ term by external fields.} \label{fig:sim}
\vspace{-0.5cm}
\end{figure}

The individual cavities are coupled together by an interaction
$$
H_{hop}=J_a(a_i^\dagger a_{i+1}+a_ia_{i+1}^\dagger)+J_b(b_i^\dagger
b_{i+1}+b_ib_{i+1}^\dagger),
$$
where $J_a,J_b\ll U$ correspond to the hopping strengths for the two different
polarizations of photons between neighbouring cavities \cite{angelakis-2006}.
The effect of the coupling can be studied by applying perturbation
theory (to the second order) to a pair of neighbouring sites, using
the formula
$$
H_{\text{eff}}=\sum_{a,b\in\{0,1\}^2}\ket{b}\bra{a}\sum_{\mu}
\frac{\bra{b}H_{hop}\proj{\mu}H_{hop}\ket{a}}{E-E_{\mu}},
$$
where $\ket{\mu}$ are all possible eigenvectors involving
 2 excitations on 1 site, and none on the other. Calculating the relevant
matrix elements in the $\ket{0}$,$\ket{1}$ basis we find the
effective interaction Hamiltonian
\begin{equation}
H_{\text{eff}}=-B_z(\identity\otimes
Z+Z\otimes\identity)- \lambda_zZ\otimes Z-\lambda_x(XX+YY)  \label{eqn:Heff}
\end{equation}
where
\begin{eqnarray}
\kappa=\frac{31}{32g}\left(J_a^2+J_b^2\right) && B_z=\frac{5}{8g}\left(J_a^2-J_b^2\right)  \nonumber\\
\lambda_z=\frac{9}{32g}\left(J_a^2+J_b^2\right)   && \lambda_x=\frac{9J_aJ_b}{16g}   \nonumber
\end{eqnarray}
and we have ignored the term $\kappa\identity$ which simply contributes a global phase. The
local magnetic fields can be manipulated by applying local Stark fields of our
own, thereby leaving an XXZ Hamiltonian where the coefficients
$\lambda_z$ and $\lambda_x$ are independently tunable (at
manufacture of the device). A comparison of the theoretical prediction and an exact diagonalization are depicted in Fig.~(\ref{fig:sim}). A degree of tunability of the Hamiltonian can be introduced at run-time by varying the detunings of the atomic transitions. However, one must remain in the regime where the detuning is small so that the perturbative expansion still holds, which restricts the range of variation.

{\em Generalised Model:} Our hopping terms, with strengths $J_a$ and
$J_b$, effectively describe transmission of photons (between
cavities) through a birefringent crystal with fast and slow axes
aligned with the directions $a$ and $b$. In an optical lattice, one
can rotate these axes by applying a Raman transition to the
tunnelling potential. In CCAs, the ability to apply this rotation is
dependant on the particular realisation under consideration. In a
setting where the cavities are connected by optical fibres, such as
fibre-coupled micro-toroidal cavities \cite{toroid}, these optical
fibres represent the birefringent material that we require, and the
optical axes ($c$ and $d$) can be aligned independently from the
directions defined by the atomic transitions ($a$ and $b$).
Moreover, the degree of birefringence ($J_a/J_b$) and the
orientation can potentially be tuned during the experiment by
applying an electric field perpendicular to the fibre, and making
use of the Kerr effect, rather than having to initialise all of
these properties at the point of manufacture. In Circuit QED and
photonic crystal realisations, however, the hopping comes directly
from the overlap of the wavefunctions of the individual
sites \cite{supercond,pbgs}, which are thus directly connected to the
$a/b$ basis, and it seems unlikely that these will support this
generalisation. In cases where this rotation can be achieved, the
two sets of axes are unitarily related,
$$
\left(\begin{array}{c} c^\dagger \\ d^\dagger \end{array}\right)
=V\left(\begin{array}{c} a^\dagger \\ b^\dagger \end{array}\right)
$$
and the
simulated Hamiltonian is changed to
$(V\otimes V)H_{\text{eff}}(V\otimes V)^{\dagger}.$
While this generates a
variety of different terms, for example $X_1Y_2+Y_1X_2$, we are unable to realise the fully anisotropic model XYZ.

\begin{figure}
\begin{center}
\includegraphics[width=0.15\textwidth]{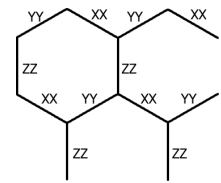}
\end{center}
\vspace{-0.5cm}
\caption{In a hexagonal lattice, with spin-1/2 located at each vertex, and the indicated couplings, topological effects arise.}
\label{fig:hexagon}
\vspace{-0.5cm}
\end{figure}

One very useful simulation that is introduced due to this  rotation
is that of the hexagonal lattice \cite{DDL03a}. At one limit, this
yields the toric code \cite{DKLP02a}, and in another region yields
non-Abelian anyons with the aid of an external magnetic field. It is
readily formed by setting $J_b=0$, which implies that $\lambda_x=0$,
and then rotating, along set directions, the remaining term $ZZ$
into $XX$ and $YY$ as required (see Fig.~\ref{fig:hexagon}).

Within the optical lattice community, the possibility of setting $\lambda_x=\lambda_z=0$ has been explored with a view to eliminating two-body
terms, so the leading order of perturbation theory gives three-body
interactions. Armed with this toolbox, one could generate many
interesting effects such as chiral terms \cite{pachos-2004-70}. In optical lattices, this
possibility is achieved by using a Feshbach resonance, such
that the collisional energies $U$ can be tuned arbitrarily. In the
present system, in order to set $\lambda_x=\lambda_z=0$, one requires $J_a=J_b=0$ i.e.~the spins are not coupled, and so three body terms cannot arise. We might hope to mimic the effect of Feshbach resonances by
introducing a detuning between the atom and the cavity, which would
serve to shift the energy levels. However, in order to maintain the
system's integrity, such a detuning should be $\Delta_{A,B}\ll g$, in
which case the shift in energies is unable to entirely cancel the
$\lambda_z$ term.

{\em Higher Spin Models:} Unlike the simple two-level dopant
considered in \cite{angelakis-2006}, changing the average number of
excitations per site influences the Hamiltonian that is simulated.
If there is an average of $S$ excitations per site, where $S$ is an
integer, then there are $S+1$ ground states, $\ket{\Psi^-_{S,n}}$,
for $n=0$ to $S$, enabling the simulation of a spin-$\half S$
particle. Again, there is an energy barrier of
$U=(2\sqrt{S}-\sqrt{S+1}-\sqrt{S-1})g\sim gS^{-3/2}$ to having any
number other than exactly $S$ excitations on each lattice site, so
the ground state is the Mott phase for small $J/U$. All of these models can simulate
a Hamiltonian of the form in Eqn.~(\ref{eqn:Heff}), except with
differing coupling coefficients, where the spin operators take on
the form of the generalised $SU(2)$ $X$, $Y$ and $Z$ rotations
respectively for the spin $\half S$. For example, with 2 excitations
per site, we realise an array of qutrits interacting through a form
described by Eqn.~(\ref{eqn:Heff}), where $X$, $Y$ and $Z$ are replaced by 
the equivalent qutrit operators,
\begin{eqnarray}
J_X=\left(\begin{array}{ccc}
0 & 1 & 0 \\
1 & 0 & 1 \\
0 & 1 & 0 \end{array}\right)/\sqrt{2}&\qquad&
J_Y=\left(\begin{array}{ccc}
0 & -i & 0 \\
i & 0 & -i \\
0 & i & 0 \end{array}\right)/\sqrt{2},  \nonumber
\end{eqnarray}
$J_Z=-i[J_X,J_Y]$ and
\begin{eqnarray}
\kappa=\frac{124\sqrt{2}}{7g}\left(J_a^2+J_b^2\right) && B_z=\frac{53}{2\sqrt{2}g}\left(J_a^2-J_b^2\right)  \nonumber\\
\lambda_z=\frac{123}{7\sqrt{2}g}\left(J_a^2+J_b^2\right)   && \lambda_x=\frac{123\sqrt{2}J_aJ_b}{7g}.    \nonumber
\end{eqnarray}
Again, further
refinements can be incorporated by implementing the polarization
rotations due to the presence of a birefringent material. If the rotation is described by the $2\times 2$ unitary,
$$
V=e^{-i\theta(n_xX+n_yY+n_zZ)},
$$
then the effective Hamiltonian is rotated by
$$
V'=e^{-i\theta(n_xJ_X+n_yJ_Y+n_zJ_Z)}.
$$
The functional form of the coupling constants for arbitrary $S$ can
be calculated, but is pathological. We note, however, that the
leading order matrix elements are $O(\sqrt{S})$. This significantly
adds to the diversity of models that can be efficiently simulated in
this simple model, just by changing the number of excitations
present in the initial state of the system.

{\em Non-integer filling:} Given that an integer number of
excitations, $S$, per lattice site describes spin-$\half S$
particles, a non-integer value of average excitations per site
potentially describes a blend of different types of particles.
Consider the general case where the filling fraction is $S+f$,
$0\leq f<1$. The minimum energy configuration is for a mixture of
particles of spin $\half S$ and $\half(S+1)$ in the ratio $(1-f):f$.
The analysis of first-order perturbation theory on $H_{hop}$,
yields, for the low energy dynamics, a swapping of the particles
between the sites, governed by the effective Hamiltonian
\begin{eqnarray}
H_{\text{eff}}\ket{\Psi^-_{S,i}}\ket{\Psi^-_{S+1,j+1}}=&&   \nonumber\\
&&\!\!\!\!\!\!\!\!\!\!\!\!\!\!\!\!\!\!\!\!\!\!\!\!\!\!\!\!\!\!\!\!\!\!\!\!\!\!\!\!\!\!\!\!\!\!\!\!\!\!\!\!\!\!\!\!\!\!\!\!\!\!\!\!\!\!\!\!\frac{(\sqrt{S}+\sqrt{S+1})^2}{4(S+1)}\left(J_a\sqrt{(i+1)(j+1)}\ket{\Psi^-_{S+1,i+1}}\ket{\Psi^-_{S,j}}\right. \nonumber\\
&&\!\!\!\!\!\!\!\!\!\!\!\!\!\!\!\!\!\!\!\!\!\!\!\!\!\!\!\!\!\!\!\!\!\!\!\!\!\!\!\!\!\!\!\!\left.+J_b\sqrt{(S-i+1)(S-j)}\ket{\Psi^-_{S+1,i}}\ket{\Psi^-_{S,j+1}}\right), \nonumber
\end{eqnarray}
which should be symmetrised for the possibility where the higher spin particle starts on the left. As already discussed, to first order, there is no interaction between particles of the same type. While we are unaware of a physical phenomena that this simulates, it completes the analysis of the system in the on-resonance case, and demonstrates the potential that coupled cavity arrays possess.

{\em Conclusions:} We have described a scheme to realize a family of
spin systems in an array of coupled cavities. By introducing a
V-configuration to the dopants, the range of nearest-neighbour
Hamiltonians that can be simulated is vastly enhanced. With an
integer average of $S$ excitations per site, we simulate
nearest-neighbour spin-$\half S$ interactions. For $S=1$, the
spin-$\half$ model allows us to reproduce the Heisenberg, XX and XXZ
models as well as those that exhibit both phase transitions and
topological features. In the case of a non-integer filling fraction,
we simulate a mixture of two particle types interacting. The
resultant strong spin-spin coupling and the individual
addressability of the separated cavity-atom systems make this
approach a promising step towards the realization of quantum
simulators for many-body spin problems. Since completing this work, we have become aware of other work which has considered the same V-system \cite{chineseV}, which just considered the case of $S=1$, recovering the same results presented here.

The results presented here for the simulation of spin-$\half S$
particles are exact, up to terms $O(J^3/g^2)$ and in the absence of
decoherence ($J=\max(J_a,J_b)$). The primary causes of decoherence
are photon loss from the cavities and spontaneous emission from the
atoms, whose rates are given by $\kappa$ and $\gamma$ respectively.
If decoherence is present in the system, our results remain valid
while the corresponding rates are dwarfed by the effective hopping
rates, which requires $\sqrt{S}J^2/g\gg
\max(\sqrt{S}\kappa,\gamma)$.

 Note that even though this scheme is not
especially robust against decoherence, such losses cause  the system
to leave the computational subspace and are thus detectable in the
final measurement steps. In order to more successfully combat
decoherence, one must utilise a scheme where the states of interest
are ground or dark states rather than excited states. Work is
progressing in that direction, with results to date requiring the use of a complex scheme employing constant external fields,
an elaborate detuning configuration and weakly coupled cavities 
\cite{jaeyoon_spins}. Nevertheless, it may be possible to find
interesting regimes within our model where quantum phenomena
persist, even in the presence of decoherence. For example, in \cite{angelakis-bose}, it is described how entanglement can persist in the steady state between a pair of noisy cavities when coupled through a third, pumped, cavity. Although this work makes no reference to how such a scheme might scale, or what information might usefully be extracted, it suggests that further investigation is warranted.
The case of non-integer filling fraction is, in fact, more robust to
decoherence because it only utilises first-order perturbation
theory, and hence we work in a regime where $g\gg
JS\gg\max(\sqrt{S}\kappa,\gamma)$. For the case of circuit QED
recently  $g/\max(\kappa,\gamma)\sim 400 $ has been
reported \cite{supercond}.

Another intriguing case to study is the atomic V-system in the
off-resonant case, and see how
the behaviour of the two different photon types mimics those of
two-species or single species spinor Bose Condensates
(see, for example, \cite{moore}), which should be different in nature to the non-integer fractional filling discussed here (it has the potential to allow particles to change type).

This work was supported by the Clare College Cambridge, the European
Union through the Integrated Project SCALA (CT-015714) and the
National Research Foundation $\&$ Ministry of Education, Singapore.

\end{document}